\documentclass[12pt]{iopart_mod}

\usepackage{amsmath}
\usepackage{amsfonts}
\usepackage{mathrsfs,cite}
\usepackage{graphicx,color} 
\newcommand{\beqs}{\begin{equation*}}
\newcommand{\beq}{\begin{equation}}

\newcommand{\eeqs}{\end{equation*}}
\newcommand{\eeq}{\end{equation}}

\newcommand{\beqas}{\begin{eqnarray*}}
\newcommand{\beqa}{\begin{eqnarray}}

\newcommand{\eeqas}{\end{eqnarray*}}
\newcommand{\eeqa}{\end{eqnarray}}

\newcommand{\eq}[2]{\begin{equation} #1 \label{#2} \end{equation}}

\newcommand{\eps}{\varepsilon}
\newcommand{\al}{\alpha}
\newcommand{\be}{\beta}

\newcommand{\de}{\delta}
\newcommand{\om}{\omega}

\newcommand{\la}{\lambda}

\newcommand{\ep}{\epsilon}

\newcommand{\Ga}{\Gamma}

\newcommand{\cR}{\mathcal R}

\newcommand{\cK}{\mathcal K}

\newcommand{\cO}{\mathcal{O}}
\newcommand{\cT}{{\cal T}}

\newcommand{\cH}{\mathcal{H}}

\newcommand{\blist}{\begin{itemize}}
\newcommand{\elist}{\end{itemize}}

\DeclareMathOperator{\extdm}{d}
\newcommand{\extd}{\extdm \!}

\providecommand{\href}[2]{#2}

\begin{document}

\title{Canonical bifurcation in higher derivative, higher spin, theories}
\author{S.~Deser, S.~Ertl and D.~Grumiller}

\address{Physics Department, Brandeis University, Waltham MA 02454 and \\
Lauritsen Laboratory, California Institute of Technology, Pasadena CA 91125\\
           Institute for Theoretical Physics,
           Vienna University of Technology,\\
           Wiedner Hauptstr.~8-10/136,
           Vienna, A-1040, Austria} 

\eads{\mailto{deser@brandeis.edu}, \mailto{sertl@hep.itp.tuwien.ac.at}, \mailto{grumil@hep.itp.tuwien.ac.at}}

\begin{abstract}
We present a non-perturbative canonical analysis of the $D=3$ quadratic-curvature, yet ghost-free, model to exemplify a  novel, ``constraint bifurcation'', effect.
Consequences include a jump in excitation count: a linearized level gauge variable is promoted to a dynamical one in the full theory. We illustrate these results with their concrete perturbative counterparts.
They are of course mutually consistent, as are perturbative findings in related models.
A geometrical interpretation in terms of propagating torsion reveals the model's relation to an (improved) version of Einstein--Weyl gravity at the linearized level.
Finally, we list some necessary conditions for triggering the bifurcation phenomenon in general interacting gauge systems.
\end{abstract}



\section{Introduction}

Canonical analysis {\`a} la Dirac is a straightforward, if sometimes labyrinthine, approach to 
counting a system's physical degrees of freedom (DoF)
in the presence of gauge symmetries and non-linear interactions. However, this approach can uncover unexpected subtleties, as already exemplified by some toy models in \cite{Henneaux:1992}. 
In this paper, we show that more physically motivated theories can also contain similar subtleties, with qualitatively important consequences.
Our focus will be on a specific self-interacting spin-2 gravity model, but similar effects may well arise in other interacting theories with higher-spin gauge symmetries, at least if some
(listed) necessary conditions are met.

The theory we shall study in detail is the truncation of $D=3$ ``NMG'' \cite{Bergshoeff:2009hq} to its pure quadratic curvature, yet ghost-free, part 
\cite{Deser:2009hb},
\eq{
I[g]= \frac{1}{16}\int \extd^3x\sqrt{-g}\, \left[G^{\mu\nu}G_{\mu\nu}-\frac12\,G^2 \right] = \frac{1}{16}\int \extd^3x\sqrt{-g}\, G_{\mu\nu}S^{\mu\nu}, \quad S_{\mu\nu} := R_{\mu\nu}-\frac14 g_{\mu\nu}R\,;
}{eq:action2}
here $G_{\mu\nu}$ is the Einstein tensor, $G$ is its trace and $S_{\mu\nu}$ is the $D=3$ 
Schouten tensor. We have set an overall dimensional constant to unity and used mostly plus signature.

This model provides a remarkable example of symmetry-breaking through the clash between its two local, conformal and coordinate, invariances. 
Their co-existence at linearized level underlies this fourth derivative metric system's ``miraculous'' transmutation into single ghost-free vector excitation, or equivalently to a propagating
torsion with non-propagating metric.
Nonlinearly, however, conformal- is necessarily sacrificed to coordinate-invariance. 

This paper is organized as follows:
In section \ref{se:2} we perform the Hamiltonian analysis. 
In section \ref{se:3} we exhibit the bifurcation mechanism and count the number of physical degrees of freedom.
In section \ref{se:4}, we first transmute the free field into a geometric model with propagating torsion, relating it to an (improved) version of Einstein--Weyl gravity, as well as to its other, Maxwell vector, avatar. We then exhibit the nonlinear obstructions and their effects, in particular, introduction of propagator-less variables.  
In section \ref{se:5} we comment on possible applications to other interacting gauge theories.

\section{Hamiltonian analysis of Schouten gravity}\label{se:2}

In keeping with our Hamiltonian approach, we use a first-order formulation of the action, 
\eq{
I[e,\om,f,\la]= \int\extd^3x \left[\frac12 \ep^{\mu\nu\rho}f^i{}_{\mu}R_{i\nu\rho}+\frac12 \ep^{\mu\nu\rho} \la^i{}_{\mu} T_{i\nu\rho}-\frac{e}{4}\left(f_{ik}f^{ik}-f^2\right)\right]\,.
}{eq:Lag1}
Roman/Greek indices are local/world. 
In form notation, the Cartan, dreibein $e^i$ (with determinant $e$) and (dualized) spin-connection $\om^i$, variables define the 
torsion and (dualized) curvature $T^i=\extd e^i+\eps^i{}_{jk}\,\om^j e^k$, $R^i=\extd\om^i+\eps^i{}_{jk}\,\om^j\om^k$.
The Lagrange-multiplier $\la^i$ ensures the on-shell torsion constraint $T^i=0$, while the auxiliary $f^i$ is essentially the same as $F^{\mu\nu}$ in \cite{Deser:2009hb} and in the Appendix A of the second reference \cite{Afshar:2011yh}. 
The first-order action \eqref{eq:Lag1} is just an Ostrogradsky auxiliary variable form of the original fourth-order action \eqref{eq:action2}, so the two are classically equivalent and
share the same excitation content. 

We now proceed to analyze \eqref{eq:Lag1} canonically, following the earlier methods of \cite{Blagojevic:2010ir,Afshar:2011yh}
, to which we refer for a more extensive discussion.
The Lagrangian's variables ($e^i{_\mu}, \om^i{_\mu},\la^i{_\mu},f^i{_\mu})$ have conjugate momenta $(\pi_i{^\mu},$ $\Pi_i{^\mu},$ $p_i{^\mu},$ $P_i{^\mu})$, leading to the primary constraints ($\approx$ means weakly equal, i.e., equal on the constraint surface):
\begin{subequations}
\begin{align}
& \phi_i{^0} :=\pi_i{^0}\approx 0 \, ,  & 
& \phi_i{^\al} :=\pi_i{^\al}-\epsilon^{0\al\be}\la_{i\be}\approx 0\, , \\  
& \Phi_i{^0} :=\Pi_i{^0}\approx 0\, ,   & 
& \Phi_i{^\al} :=\Pi_i{^\al}-\epsilon^{0\al\be} f_{i\be}\approx 0\,,  \\
& p_i{^\mu} \approx 0\, , && P_i{^\mu}\approx 0\, . \label{prime}
\end{align}
\end{subequations}
The constraints $(\phi_i{^\al},\Phi_i{^\al},p_i{^\al},P_i{^\al})$ are second class.
It will be useful to define the linear combination 
\eq{
\tilde\phi_i{}^0:=\phi_i{}^0 + f_i{}^kP_k{}^0 + \la_i{}^k p_k{}^0\, .
}{eq:lincom}
Eliminating the momenta $(\pi_i{^\al},\Pi_i{^\al},p_i{^\al},P_i{^\al})$ leads to a partly reduced phase space, in which the nontrivial Dirac 
brackets are given by
\begin{align}\label{eq:nontrivialdirac}
&\{e^i{_\al},\la^j{_\be}\}=\eta^{ij}\epsilon_{0\al\be}\, ,  &  
&\{\om^i{_\al},f^j{_\be}\}=\eta^{ij}\epsilon_{0\al\be}\,. 
\end{align}
The remaining Dirac brackets are the same as the corresponding Poisson brackets. 
The canonical Hamiltonian $\cH_c$ can be written as a sum over secondary constraints
\eq{
\cH_c := e^i{}_0{\cal H}_i+\om^i{}_0{\cal K}_i+f^i{_0}{\cal R}_i
      +\la^i{_0}\cT_i 
}{eq:Hcannew}
up to boundary terms. Here,
\begin{subequations}
\begin{align}
\cH_i &:=-\epsilon^{0\al\be}{\cal D}_\al\la_{i\be} +e \left(e_i{}^0 {\cal V}_K-\frac12\,f_{ik}(f^{k0}-f e^{k0})\right)\approx 0\, , \\
\cK_i &:=-\epsilon^{0\al\be}\Big({\cal D}_\al f_{i\be} 
        +\eps_{ijk}e^j{}_\al \la^k{}_\be\Big) \approx 0\,,           \\
\cR_i &:=-\frac{1}{2}\epsilon^{0\al\be}R_{i\al\be} + \frac12\,e \left(f_i{}^0-fe_i{}^0\right)\approx 0   \,,          \\
\cT_i &:=-\frac{1}{2}\epsilon^{0\al\be}T_{i\al\be} \approx 0 \,, 
\end{align}
\end{subequations}
and
\eq{
{\cal V}_K :=\frac14\left(f_{ik}f^{ij}-f^2\right)\,.
}{eq:VK}
The consistency conditions of the secondary constraints lead to the ternary constraints $\theta_{\mu\nu}\approx0\approx\psi_{\mu\nu}$ that establish the symmetry of the auxiliary fields.
Similarly, the consistency conditions of the ternary constraints yield quaternary constraints
\eq{
\chi=\la\approx 0, \qquad\qquad \varphi=f+\frac{1}{2}{\cal V}_K\approx 0\,.
}{}
At this stage the constraint procedure fortunately ends: no further constraints are generated through consistency conditions.

\section{Counting degrees of freedom: Bifurcation}\label{se:3}

Having found all constraints in the previous section, we establish their first/second class properties in order to count the number of physical DoF.
We summarize the main result in tables \ref{tab:1} and \ref{tab:2}.
The fact that we have two different tables is a consequence of the advertised subtlety we called ``bifurcation'' and now explain.

The consistency condition of the quaternary constraint $\varphi$ leads to the following Dirac bracket
\begin{equation}\label{eq:quatcons}
\{\varphi,\,\cH_T\} =\frac{1}{4}\left(f^{\mu\nu}-f g^{\mu\nu}\right)\,z_{\mu\nu} \approx 0\, , 
\end{equation}
where $\cH_T$ is the total Hamiltonian
\eq{
\cH_T := \cH_c + u^i{}_0\tilde\phi_i{}^0 + v^i{}_0\Phi_i{}^0 + w^i{}_0 p_i{}^0 + z^i{}_0P_i{}^0
}{eq:HT}
and ($u$, $v$, $w$, $z$) are Lagrange-multipliers of primary constraints.
All components $w_{\mu\nu}$ and $z_{\mu\nu}$ are determined at this stage (and hence the corresponding constraints are second class), except for the Lagrange multiplier $z_{00}$ that multiplies the primary constraint $P^{00}$.
The consistency conditions \eqref{eq:quatcons} allows us to determine the remaining Lagrange multiplier $z_{00}$, unless the condition
\eq{
f^{00} = f g^{00} 
}{eq:condition}
holds. The resulting bifurcation consists in the following: if condition \eqref{eq:condition} does/not hold, the quaternary constraint $\varphi$ and the primary constraint $P^{00}$ are both first/second class.

\begin{table}[h]
\begin{center}
\doublerulesep 1.8pt
\begin{tabular}{||l|l|l||}
                                                      \hline\hline
\rule{0pt}{12pt}
&~First class \phantom{x}&~Second class \phantom{x} \\
                                                      \hline
\rule[-1pt]{0pt}{15pt}
\phantom{x}Primary &~$\tilde\phi_i{}^0,\Phi_i{^0}$
            &~$p_i{^0},P_i{^0}$   \\
                                                      \hline
\rule[-1pt]{0pt}{15pt}
\phantom{x}Secondary\phantom{x} &~${\bar {\cal H}}_i,{\bar {\cal K}}_i$
           &~$\cT_i,{\hat{\cal R}}'_i$       \\
                                                      \hline
\rule[-1pt]{0pt}{15pt}
\phantom{x}Ternary\phantom{x}
                  & &~$\theta_{0\be},\theta_{\al\be},\psi_{0\be},\psi_{\al\be}$ \\
                                                      \hline
\rule[-1pt]{0pt}{15pt}
\phantom{x}Quaternary\phantom{x}
                 &  &~$\chi, \varphi$  \\
                                                      \hline\hline
\end{tabular}
\end{center}
\caption{Classification of constraints in the partly reduced phase space absent the constraint \eqref{eq:condition}.}
\label{tab:1}
\end{table}

\begin{table}[h]
\begin{center}
\doublerulesep 1.8pt
\begin{tabular}{||l|l|l||}
                                                      \hline\hline
\rule{0pt}{12pt}
&~First class \phantom{x}&~Second class \phantom{x} \\
                                                      \hline
\rule[-1pt]{0pt}{15pt}
\phantom{x}Primary &~$\tilde\phi_i{^0},\Phi_i{^0}$, $P^{00}$
            &~$p_i{^0}$, remaining $P_i{^0}$   \\
                                                      \hline
\rule[-1pt]{0pt}{15pt}
\phantom{x}Secondary\phantom{x} &~${\bar {\cal H}}_i,{\bar {\cal K}}_i$
           &~$\cT_i,{\hat{\cal R}}'_i$       \\
                                                      \hline
\rule[-1pt]{0pt}{15pt}
\phantom{x}Ternary\phantom{x}
                  & &~$\theta_{0\be},\theta_{\al\be},\psi_{0\be},\psi_{\al\be}$ \\
                                                      \hline
\rule[-1pt]{0pt}{15pt}
\phantom{x}Quaternary\phantom{x}
                 & $\varphi$ &~$\chi$  \\
                                                      \hline\hline
\end{tabular}
\end{center}
\caption{Classification of constraints in the partly reduced phase space with the constraint \eqref{eq:condition} present.}
\label{tab:2}
\end{table}

According to table \ref{tab:1}, i.e., when \eqref{eq:condition} is absent, we have a 48-dimensional phase space with 12 first class and 20 second class constraints.
Consequently, the theory 
in general exhibits two local physical DoFs, namely the massive bulk gravitons. 
However, if the condition \eqref{eq:condition} holds, then table \ref{tab:2} states that there are 14 first class constraints and 18 second class constraints.
In that case the number of local physical DoFs is reduced to one.


Translating condition \eqref{eq:condition} into metric form, by using $f_{\mu\nu}=2S_{\mu\nu}$ and $f=\frac12 R$,
one finds that it holds automatically if the metric is a solution of the Einstein equations 
\eq{
 R^{\mu\nu}-\frac12 g^{\mu\nu}R=0 \quad\rightarrow\quad f^{\mu\nu}=g^{\mu\nu}f \,.
}{}
This observation concurs with the fact that the linearized theory around Einstein solutions has one additional gauge symmetry due to partial masslessness \cite{Deser:2001us}.
Again, this symmetry enhancement results from the fact that the constraint $P^{00}$ becomes first class, leading to the additional gauge symmetry.
However, the latter is an artifact of linearization, being broken in the full nonlinear theory, as we have shown above.

\section{Conformal versus coordinate invariance}\label{se:4}

In this section we provide an alternative derivation of the loss of gauge-invariance beyond linearization in a straightforward perturbative approach.
This sheds additional light on the bifurcation mechanism and exhibits it as a clash between diffeomorphism and conformal invariance. 
Moreover, the perturbative analysis can be useful particularly for possible higher-rank/higher-spin generalizations, where a full canonical analysis is often less accessible than a perturbative one.

We work with the second order, ``Ostrogradski'' action [equivalent to \eqref{eq:action2}], using auxiliary, symmetric tensor density, variables $f^{\mu\nu}$:
\eq{
I [g,\,f]=  \frac{1}{4}\int \extd^3x\; \Big\{G_{\mu\nu}(g)\, f^{\mu\nu} - \frac12 \big[f_{\mu\nu}^2 -(\textrm{Tr}\, f)^2\big]/ \sqrt{-g}\Big\}  
}{eq:saction2}
where we have omitted the contracting metrics.
[Completing squares and integrating out $f$ recovers (\ref{eq:action2}).] We will study (\ref{eq:saction2}), initially at linear, then full non-linear, metric levels. 
The linearization of (\ref{eq:action2}) is manifestly (linear) diffeo-invariant, while use of the Bianchi identity easily confirms its conformal invariance, under   
\begin{align}\label{eq:eq3}
\de h_{\mu\nu} = -2\eta_{\mu\nu}\al,\quad\de G_{\mu\nu} = (\partial_{\mu}\partial_{\nu}-g_{\mu\nu}\square) \al\,,
\end{align}
with our convention $R_{\mu\nu}=\partial_\la \Ga^\la{}_{\nu\mu}-\partial_\nu\Ga^\la{}_{\la\mu} + \dots$.
Both invariances also hold in (\ref{eq:saction2}), of course, with $f$ transforming as a (linear) diffeo tensor and conformally like the Schouten tensor: $\de f_{\mu\nu} = \partial_{\mu}\partial_{\nu}\, \al$. 

Now we count DoF: Varying $h_{\mu\nu}$ yields $G_{\mu\nu}(f)=0$, where $G$ is the usual linear Einstein operator. In $D=3$, there are no Einstein excitations (Riemann and Ricci being equivalent), so 
$f_{\mu\nu}$ is a pure gauge ``metric''. Varying $f$, we learn that $h_{\mu\nu}$ obeys the Einstein equation with linear source, so the general solution is:
\begin{align}
f_{\mu\nu} &= \partial_{\mu} A_{\nu} + \partial_{\nu} A_{\mu}\,,\quad\; G_{\mu\nu}(h) = f_{\mu\nu} -\eta_{\mu\nu} f\,. \label{eq:sol4}
\end{align}
Inserting -- legally -- (\ref{eq:sol4}) into (\ref{eq:saction2}), we note first that its $G(h)\, f= 2 G^{\mu\nu} \partial_{\mu} A_{\nu}$ term vanishes 
by the Bianchi identities upon part integration, leaving the quadratic $f$-terms: these precisely combine into the promised reduced one-DoF Maxwell action (second reference in \cite{Bergshoeff:2009hq}):
\eq{
I[h,\, f] \rightarrow \; -\frac14 \int \extd^3x\; F_{\mu\nu}^2, \quad  F_{\mu\nu}:=(\partial_{\mu} A_{\nu} - \partial_{\nu} A_{\mu})\,;
}{eq:eq5}
it is invariant under $\de A_{\mu}= \frac{1}{2}\partial_{\mu} \al$. That $\al$ is indeed our conformal transformation parameter   
\begin{align}
\delta f_{\mu\nu} =  \partial_\mu\partial_\nu\alpha 
\end{align}
is then verified by (\ref{eq:eq3},\ref{eq:sol4}).
The above set of field equations is consistent with both underlying invariances; for example, upon taking the divergence of Einstein equation in (\ref{eq:sol4}), its left side vanishes by the Bianchi identity, 
while the divergence/conservation of its (symmetric) right side matter source also does, being proportional to the latters's field, i.e., Maxwell's, equations. 
In this connection, note that there is no ``spin-loss'' paradox in the above tensor-to-vector transmutation because all
massless fields in $D=3$ are necessarily spinless \cite{Binegar:1981gv,Deser:1991mw}, at least in flat space; it would be useful to learn if this persists in (A)dS.

The above, conformal-to-gauge transmutation should not be confused with a separate, surprising \cite{JP:2011}, $D=3$ conformal invariance 
enjoyed by Maxwell, by virtue of its further transmutability into a scalar. We provide a concise derivation, emphasizing the nonlocal 
(as usual with such transmutations) price involved: The first order Maxwell action is  
\eq{
I [F,A] = - \frac12 \int \extd^3x \big[F^{\mu\nu} (\partial_{\mu} A_{\nu} - \partial_{\nu} A_{\mu}) - \frac12 F_{\mu\nu}^2\big]\,,
}{eq:A}
where $F^{\mu\nu}$ and $A_{\mu}$ are independent variables. Varying $A_{\mu}$ gives $\partial_{\nu} F^{\mu\nu}=0$, 
whose general solution is $F^{\mu\nu}=\eps^{\mu\nu\al} \partial_\al S$, the scalar $S$ having dimension of $A_\mu$; $F$'s dual vector $\ast F_\mu$ is just $\partial_\mu S$, 
a relation whose highly non-local inverse we also note,
\eq{
S(\ast F)= \square^{-1} \partial_{\mu} \ast F^{\mu}\,. 
}{eq:B}
Replacing $\ast F$ by $\partial S$ in \eqref{eq:A} immediately yields the free scalar action, 
$I[F,A] \rightarrow I[S] = - \frac12 \int \extd^3x (\partial S)^2$. To cancel the scalar stress-tensor's trace, add to its minimal $T_{\mu\nu}(S)$
the usual identically conserved (hence not affecting the Poincare generators) improvement \cite{Callan:1970ze} term, $\de T_{\mu\nu}\sim -\frac14 \big(\partial_{\mu} \partial_{\nu}-\eta_{\mu\nu} \square\big) S^2$.
Its on-shell $\textrm{Tr}(\partial S)^2$ cancels that of $T_{\mu\nu}$.
The above procedure is nonlocal in $F_{\mu\nu}$, inherited from $S(\ast F)$ in \eqref{eq:B}. But one cannot improve directly at the vector level either. Any such 
attempt is doomed from the start:  it is easily verified  that there is no identically conserved $\de T \sim dd(AA)$ -- as required by dimension -- 
that is even on-shell gauge-invariant, so one cannot cancel the Maxwell stress-tensor's trace, $\sim \frac14 F_{\mu\nu}^2$.  Nor is any other local choice 
available, since it must, just by dimensions, depend on $AA$ rather than on the field strengths $FF$; the scalar itself is the only gauge-invariant, dimensionally correct,  
but nonlocal, option. [We have not attempted to find a direct descent from \eqref{eq:saction2} to the scalar action.]

In canonical terms, the associated DoF-reducing Maxwell gauge constraint is the standard ($A_0\,\nabla\!\cdot {\bf E}$), leaving
a $D=2+1$ photon with just one transverse DoF, whose sign is fixed by that chosen for (\ref{eq:action2}) or (\ref{eq:saction2}). 
More explicitly, the Maxwell action's kinetic term is ($-{\bf{E}}^T\!\cdot\!\dot{\bf{A}}^T-{\bf E}^L\!\cdot\!\dot{\bf A}^L$), in terms of the spatial
transverse-longitudinal orthogonal vector decomposition $V_i=\eps_i{}^j \partial_j V +\partial_i v$, which commutes with time-derivatives.
The longitudinal excitation is removed by the Gauss constraint $(A_0 \nabla\!\cdot {\bf E})$ that enforces ${\bf E}^L=0$. Instead, the cubic correction (\ref{eq:eq6}) contains terms quadratic in $A_0$, thereby replacing this constraint
by an irrelevant (because it integrates out) perfect square plus (cubic) terms that depend on $(E^i, A_i)$, so the longitudinal DoF are reinstated. 
(We have checked that cubic terms $ \sim h_{00} (\dot{A_0})^2$ are absent though.)
Not having studied the Hamiltonian in detail, we cannot assert that it is
no longer bounded below, but that seems likely for any cubic: one would have to include quartic corrections for a meaningful conclusion, though that is not very relevant any more.
This paradoxical second order form of a fourth order action is explained by the original metric propagator indeed being \cite{Deser:2009hb} $ \sim (\nabla^2 \;\square)^{-1}$, thus agreeing with $f$ having
dimension of curvature, $f\sim ddh$, hence $A\sim dh$.
[Actually, our method also applies, with appropriate dimensional numerical
differences, to the $D=4$ Weyl action, which can also be written (modulo
coefficient differences in Schouten) in the form \eqref{eq:saction2}. The $D=3$ argument is
unchanged, in particular that $h$ and $F$ each obey the Einstein equation, so
each has 2, rather than $D=3$'s 0, DoF. The $A_{\mu}$, coordinate vector gauge,
part of $f$ still represents a (now 2 DoF) photon \cite{Deser:2009hb} (see also \cite{Deser:1980fy}).
Again, there is no spin-transmutation paradox (despite the presence of spin $D=4$), because the propagator of
a fourth derivative order theory can (and usually does) include lower-spin poles.]

Our linearized system has a second, more general geometrical interpretation: it represents a propagating vector torsion and contortion, but a non-propagating (because $D=3$) metric. 
Indeed, this is already suggested by the Einstein equation in \eqref{eq:sol4}. We recall that one may extend Riemannian geometries by extending its metric affinity
$\Ga^{\al}{}_{\mu\nu}(h)$ to include an antisymmetric torsion, $T^{\al}{}_{[\mu\nu]}$, and a contortion, whose symmetric part we denote by $K^{\al}{}_{(\mu\nu)}$ [we define (anti-)symmetrization with a factor 1/2]. For our system, the choices are
\eq{
T^\al{}_{\mu\nu} = b\; \de^\al_{[\mu} A_{\nu]}\,, \qquad K^\al{}_{(\mu\nu)} = c\; \de^\al_{(\mu} A_{\nu)} + d \;\eta_{\mu\nu} A^\al\,.
}{eq:geomint}
The parameter choice $d+2c=0$ kills the Ricci tensor's antisymmetric part.
The further choices $b=3$ and $c=1$ 
make the vanishing of the Einstein tensor, but now of the full connection, $G(\Ga_{\textrm{tot}})=0$, $\Ga_{\textrm{tot}}{}^{\al}{}_{\mu\nu}:= \Ga^{\al}{}_{\mu\nu}(h)+T^{\al}{}_{\mu\nu}(A)+K^{\al}{}_{(\mu\nu)}(A)$,
coincide with the source-ful Einstein equation of \eqref{eq:sol4}. The divergence of the right hand side of that equation is proportional to the Maxwell operator, whose vanishing is then assured by consistency with the metric Bianchi identity. Note that this
total connection 
$\Ga_{\textrm{tot}}$ is metric compatible in the Einstein--Weyl sense. 
\eq{
(\nabla_\mu(h) - 2 A_\mu )\, g_{\al\be} = 0 
}{eq:angelinajolie}
In this interpretation the Maxwell field 
is recognized as the Weyl potential, up to a factor and the presence of torsion, see e.g.~\cite{LeBrun:1999}. We stress that the inclusion of torsion leads to an improvement of the usual torsionless Einstein--Weyl connection, since our Ricci tensor is symmetric, while the Einstein--Weyl Ricci-tensor in general is not. More explicitly, for Einstein--Weyl we have $b=0$ and $c=-2d$; note that metric-compatibility in the Einstein--Weyl sense is guaranteed if $b+c+2d=0$. Interestingly, the tracefree part of the right Eq.~\eqref{eq:sol4} coincides precisely with the Einstein--Weyl equations (up to terms quadratic in the Weyl potential), 
see \cite{LeBrun:1999,Grumiller:2007gm} and Refs.~therein. 
This provides yet-another-way to see why conformal invariance is broken cubically: the Einstein--Weyl equations are
Weyl-invariant, but they coincide with our equations only at the linearized level, not non-linearly.

We now show explicitly how enforcing diffeo- destroys conformal- invariance beyond linear order, as it must since the full action (\ref{eq:action2}) involves the factor $(\sqrt{-g} g^{\mu\nu} g^{\al\be})\sim \cO(g^{-1/2})$, 
rather than $\cO(g^0)$. We need only consider the first, cubic, deviation, where the effect will be manifested as loss of Maxwell gauge invariance. 
Returning to (\ref{eq:saction2}) 
and inserting the linearized values \eqref{eq:sol4}, 
we find that the cubic action reduces, schematically, to the ``bare $A_{\mu}$-form''
\eq{
I^3[h_{\mu\nu},\,A_\mu] \sim \frac14\int \extd^3x\,\big[2 G_2^{\mu\nu}(h)\partial_\mu A_\nu + h_{\mu\nu} Q^{\mu\nu}(A)\big] \,,    
}{eq:eq6}
where $Q^{\mu\nu}$ (and also its integral) is an (irreducibly) gauge {\em variant} quantity bi-linear in the gauge potential $A_\mu$.\footnote{We obtain $Q^{\mu\nu} = A_{\al}\partial_{\be}\big(f^{\mu\nu}\eta^{\al\be}-\eta^{\mu\nu}f^{\al\be}\big)-\frac14\eta^{\mu\nu}\big(f_{\al\be}f^{\al\be}+f^2\big)+\frac12 ff^{\mu\nu}-f^{\mu\al}F_{\al}{}^{\nu}$.} 
This manifest loss of Maxwell gauge invariance means of course, loss of conformal invariance. 
A concrete, ``no-calculation'' realization of invariance loss is now easy. Consider first the vacuum state in the gauge $g_{\mu\nu}=\eta_{\mu\nu} $, $ A_{\mu}=0 $: all linearized and cubic terms manifestly vanish.
Now gauge-vary this ``null'' configuration, with local parameter 
 $\alpha= x^{\mu} x_{\mu}$, where $x^{\mu}$ are the Minkowski coordinates. Then $A_{\mu} = x_{\mu}, \; f_{\mu\nu} = 2\eta_{\mu\nu},\;  f = 6, \; h_{\mu\nu}\, \propto\, x^2 \, \eta_{\mu\nu}$: This means $A_{\mu}$ is a conformal Killing vector of the flat background. 
The (gauge invariant) quadratic action still vanishes of course. Instead,
$I^3 \propto \int \eta_{\mu\nu} Q^{\mu\nu}$.
This, manifestly constant trace, $\eta_{\mu\nu} Q^{\mu\nu}$, is easily found to be non-zero; hence the cubic action is a non-invariant, proportional (in this gauge) to the vector ``mass'' term 
$I^3 \propto  x^{\mu} x_{\mu} = m^2 A^{\mu} A_{\mu} \neq 0$, already at vacuum.

We conclude that the simple one-DoF content of the linearized level is indeed lost here, with the implied consequent presence of additional, ghost, modes that always plagues generic quadratic actions, albeit without
propagators of their own. 
                                                                                                                                                                                             

\section{Conclusions}\label{se:5}

We have studied a novel, higher-order, clash between two local invariances that characterize dynamical quadratic curvature Schouten Schouten-ghost-free $D=3$ gravity, at linear order. After performing a canonical analysis and exhibiting a bifurcation
in the constraint analysis that leads to different counts of DoFs, we considered the theory perturbatively, providing a transparent derivation of the free theory's one-DoF, second order, character: our geometrical, torsionful, representation complementing an earlier vectorial transmutation. We then traced the unavoidable breaking, already at cubic level, of the model's conformal symmetry by its nonlinear diffeo-invariant, but conformal-gauge dependent, completion. 
The culprits were the dynamical (rather than Minkowski) metrics that contract indices beyond lowest order. 

The above symmetry-breaking raises one, also novel, field-theoretic puzzle posed by this otherwise consistent model: Its propagator depends on fewer variables than do its vertices.
How does one calculate (at least perturbatively) either classically or at quantum loop level? The new variables, having no lines of their own, can only lie on open, but presumably not on closed loop lines,
 yet they are not external fields either.
We have also checked the consistency of our perturbative analysis with a non-perturbative canonical analysis along the lines of \cite{Afshar:2011yh}, 
with perfect agreement.
  A (vaguely) similar situation occurs in topologically massive gravity, whose metric's propagator is not uniformly of either second or third derivative order, since its third derivative Cotton 
contribution is independent of the metric's conformal factor, one that is present in its Einstein, second order term. That problem is not one of principle, however -- all components have proper propagators, 
just ones of different momentum order -- and can only affect topologically massive gravity's UV behavior \cite{Deser:1990bj}. Rather, the nearest analog is perhaps massive $D=4$ Einstein gravity with a ``wrong''
explicit mass term a la \cite{Boulware:1973my}; however that symmetry-breaking pathology is inserted by hand, rather than, as in our model, from enforcing a greater invariance! Clearly, some intriguing unsolved directions remain.

The general non-perturbative conditions of the canonical analysis we encountered here are a non-perturbative extension and here agree with the perturbative results.
Moreover, a similar effect was observed in generalized massive gravity, see second reference in \cite{Afshar:2011yh}.
This suggests that the bifurcation effect we described here is not just confined to our specific model \eqref{eq:action2}, but 
can be present in more general interacting higher-rank/higher-spin gauge theories as well as, perhaps, in vector models, see \cite{Jackiw:1997jga} and Refs.~therein.

We conclude with a list of ingredients necessary, but not always sufficient, for the emergence of bifurcation.
\begin{enumerate}
 \item Higher (at least spin 2) gauge symmetries 
 \item Non-linear (at least cubic) interactions 
 \item The linearized theory must have additional (linearized) gauge symmetries such as conformal/Weyl symmetry
 \item The linearized gauge symmetry must be broken in the interacting theory
\end{enumerate}
It would be interesting to construct explicit examples of bifurcation in $D>3$ theories with spins higher than 2, and/or towers thereof.

\section*{Acknowledgments}

We thank H.~Afshar, J.~Franklin, O.~Hohm, R.~Jackiw and P.~Townsend for useful discussions.
DG thanks Caltech and KITP for hospitality during the KITP program `Bits, Branes, Black Holes' while part of this work was begun.
SD was supported in part by NSF PHY-1064302 and DOE DE-FG02-164 92ER40701 grants. 
SE and DG are supported by the START project Y435-N16 of the Austrian Science Fund (FWF) and by the FWF project P21927-N16.
This research was supported in part by the National Science Foundation under Grant No. NSF PHY11-25915.

\paragraph{Note added:} 
As this work was completed, and simultaneous with our submission of the perturbative analysis (essentially section \ref{se:4}), the closely related posting \cite{Hohm:2012vh} appeared.
It discusses the Chern-Simons and other extensions of the present model. Where they overlap, 
our conclusions agree; indeed the models of \cite{Hohm:2012vh} provide further evidence for our conjecture that the bifurcation we encountered here is a more generic feature of interacting higher-rank/higher-spin theories.



\section*{References}


\begingroup\raggedright\endgroup

\end{document}